\documentstyle[aps,twocolumn,epsfig,rmp]{revtex}
\def\lap{\hbox{~{\lower -2.5pt\hbox{$<$}}\hskip -8pt\raise
-3.5pt\hbox{$\sim$}}}
\def\gap{\hbox{~{\lower -2.5pt\hbox{$>$}}\hskip -8pt\raise
-3.5pt\hbox{$\sim$}}}
\def\apg{\hbox{{\raise -2.5pt\hbox{$>$}}\hskip -8pt\lower -
2.5pt\hbox{$\sim$}}}
\def\apl{\hbox{{\raise -2.5pt\hbox{$<$}}\hskip -8pt\lower -
2.5pt\hbox{$\sim$}}}
\def\A{{\cal A}}
\def\S{{\cal S}}
\def\s{ \{|A_k\rangle\} }
\def\a{{\alpha}}
\def\b{{\beta}}
\def\Zs{{0_{\S}}}
\def\Js{{1_{\S}}}
\def\Za{{0_{\cal A}}}
\def\Ja{{1_{\cal A}}}
\def\0d{{0_{\cal D}}}
\def\1d{{1_{\cal D}}}
\begin{document}
\bigskip
\title {QUANTUM DISCORD AND MAXWELL'S DEMONS}

\author{Wojciech Hubert Zurek}
\address{Theory Division, \\
T-6, MS B288, LANL\\
Los Alamos, New Mexico 87545}
\maketitle
\bigskip
\centerline{\it Abstract}
\medskip
\noindent {\it Quantum discord was proposed as an information theoretic measure
of the ``quantumness'' of correlations. I show that discord determines
the difference between the efficiency of quantum and classical Maxwell's
demons -- that is, entities that can or cannot measure nonlocal observables
or carry out conditional quantum operations -- in extracting work from 
collections of correlated quantum systems.}

\bigskip

\section{Demons}

Maxwell's demon$^1$ was introduced to explore the role of information
and, more generally, to investigate the place of ``intelligent observers"
in physics. In modern discussions of the subject$^2$ ``intelligence'' is often
regarded as predicated upon or even synonymous with the information processing
ability -- with computing. Thus, Maxwell's demon is frequently modeled by
a Turing machine -- a classical computer -- endowed with the ability
to measure and act depending on the outcome. The role of a demon is
to implement an appropriate conditional dynamics -- to react to the state
of the system as revealed through its correlation with the state of
the apparatus. 

It is now known that quantum logic -- i.e., logic
employed by quantum computers -- is in some applications more powerful
than its classical counterpart. It is therefore intriguing to enquire whether
a quantum demon -- an entity that can measure non-local states and implement
quantum conditional operations -- could be more efficient than a classical one.
I show that quantum demons can extract more work than classical
demons from correlations between quantum systems, and that the difference
is given by the {\it quantum discord}, a recently introduced$^{3-5}$
measure of the ``quantumness'' of correlations.

Maxwell's demon sets up a useful conceptual framework that provides an
operational interpretation of discord. The role played by the quantum demon --
carrying out conditional quantum operations on pairs of systems -- could be
also fulfilled the classical device that can outright measure non-local
quantum observables. This is especially apparent in section IV where we
alternate between the quantum and classical demon on one hand, and Alice
and Bob on the other. The real point of employing demons is to draw attention
to the thermodynamic (and information - theoretic) costs of various 
operations and -- in a sense -- to hold Alice and Bob accountable for
their thermodynamic expenditures which are usually simply ignored.

\medskip
\section{Discord}

Quantum discord$^{3-5}$ is the difference between two classically identical
formulae that measure the information content of a pair of quantum systems.
Several closely related variants can be obtained starting from the original
definition$^3$ given in terms of mutual information$^6$. Mutual information
quantifies the strength of correlations between, say,
the apparatus $\A$ and the system $\S$:
$$ I(\S:\A) = H(\S) + H(\A)  - H(\S, \A)  \eqno(1)$$
It measures the difference between the missing information about the
two objects
when they are taken separately, $H(\S) + H(\A)$, and jointly, $H(\S,\A)$
(see Fig. 1). In the extreme case $\S$ and $\A$ may be identical -- e.g.,
copies of the same book, or a state of the apparatus pointer $\A$ after
a perfect but as yet unread measurement of $\S$. Then the joint entropy
$H(\S, \A)$ is equal to $H(\A) = H (\S)$, so $I(\S: \A) = H(\A)$. By contrast,
when the two objects are not correlated, $H(\S,\A) = H(\S) + H(\A)$, and
$I(\S:\A)=0$.

The other formula for mutual information employs classical identity
for joint entropy$^6$:
$$ H(\S, \A) = H(\A) + H(\S|\A) = H(\S) + H(\A | \S) \eqno(2)$$
Above $H(\S|\A)$ is the conditional entropy -- e.g., the measure of the lack
of information about the state of ${\cal S}$ given the state of ${\cal A}$.
Substituting this in Eq. (1) leads to an asymmetric looking formula for
mutual information:
$$ J_{\cal A}(\S:\A) = H(\S) + H(\A) - [ H(\A) + H(\S|\A) ] \eqno(3)$$
We have refrained from carrying out the obvious cancellation above that would
have yielded $J_{\cal A}(\S:\A) = H(\S) - H(\S|\A)$ for a reason that will
be soon apparent.

Discord is defined as:
$$\delta(\S|\A) = I(\S:\A) - J_{\cal A}(\S:\A) =
[ H(\A) + H(\S|\A) ] - H(\S, \A) \eqno(4)$$
Classically, discord disappears as a consequence of Eq. (2) --
information about
a collection of classical objects can be acquired one object at a time. In
quantum theory, however, measurements can modify quantum state. Thus, in order
to properly define conditional entropy one must specify how the apparatus is
``interrogated'' about $\S$:
After a measurement of the observable with eigenstates $\s$ observer's own
description of the pair is the conditional density matrix:
$$\rho_{\S\A|A_k\rangle}= \rho_{\S|A_k\rangle} \otimes |A_k\rangle\langle A_k|
\eqno(5)$$
Given an outcome $|A_k\rangle$, he will attribute $\rho_{\S|A_k\rangle}
=\langle A_k|\rho_{\cal SA} |A_k\rangle/ p_{\cal A}(k)$  to $\S$
with the probability $p_{\A}(k) = Tr \langle A_k | \rho_{\S\A} | A_k \rangle $.
Even for an outsider (who has not yet found out the outcome) post-measurement
density matrix $\rho_{\S\A}'$ usually differs from the pre-measurement
$\rho_{\S\A}$. This outsider state of knowledge should be contrasted with
the viewpoint of the insider who made the measurement: Insider knows that
the apparatus is in the state $|A_k\rangle$. Outsider does not, so he obtains
his post-measurement $\rho_{\S\A}'$ by averaging over the outcomes:
$$ \rho_{\S\A}' = \sum_k p_{\A}(k) \rho_{\S| A_k\rangle} \otimes
|A_k \rangle \langle A_k| \eqno(6) $$
Outsider's description of the pair is unaffected by the insiders
measurements --
$ \rho_{\S\A}' =  \rho_{\S\A}$ --  only when the measured observable commutes
with $\rho_{\S\A}$. We shall find outsider viewpoint useful because
it represents a statistical ensemble of all possible outcomes.

In quantum physics one definition denoted by $J_{\A}(\S:\A_{\s})$ is
the {\it locally accessible mutual information}. It uses Eq. (3) with
the joint entropy given by:
$$ H_{\A}(\S, \A_{\s}) = [H(\A) + H(\S|\A)]_{\s} \eqno(7)$$
where $\s$ is the eigenbasis of the to-be-measured observable of the apparatus.
Another acceptable and completely quantum definition of $I(\S:\A)$
relies on the
the von Neumann entropy of the density matrix $\rho_{\S\A}$ describing
the joint state. Then:
$$ H(\S, \A) = - Tr \rho_{\S \A} \lg \rho_{\S \A}
= - \sum_l p_{\S\A}(l) \lg p_{\S\A}(l) \eqno(8)$$
where probabilities $p_{\S\A}(l)$ are the eigenvalues of $\rho_{\S\A}$
that describes the correlated pair. These
eigenvalues always exist, but in general correspond to entangled quantum states
$|\psi_{\S\A}(l)\rangle$
in the joint Hilbert space of $\S$ and $\A$. Such states cannot be found out
through sequences of local measurements starting with just one subsystem of
the pair -- say, $\A$. This fundamental difference between the quantum
and the classical realm (where such ``piecewise'' investigation is always
possible and need not disturb the state of the pair) is responsible for
non-zero discord.

A simple example of this situation is a perfectly entangled Bell state:
$$ |\psi_{\S\A} \rangle = (|\Zs \Za\rangle + |\Js \Ja\rangle)/\sqrt 2
\eqno(9a) $$
Clearly, $\rho_{\S\A} = |\psi_{\S\A} \rangle \langle \psi_{\S\A} |$ is pure
-- the pair is in the state $|\psi_{\S\A} \rangle$. Hence, in accord
with Eq. (8), $H(\S,\A) = 0$.
On the other hand, $\rho_{\A(\S)}=Tr_{\S(\A)} \rho_{\S\A}= {\bf 1_{\A(\S)}}/2$,
where ${\bf 1}$ is the unit matrix in the appropriate Hilbert space, so that
$H(\A) = H(\S) = 1$. Consequently, $I(\S:\A)=2$, but the asymmetric mutual
information is $J_{\A}(\S:\A)=1$. This is because the joint information
$H_{\A}(\S,\A_{\s})$ defined with reference to any measurement on a $\A$,
Eq. (5), is a sum of $H(\A) = 1$ and $H(\S|\A)=0$. In our example both of these
quantities are independent of the basis
because of the symmetry of Bell states.

Readers are invited to verify that a classical correlation in:
$$ \rho_{\S\A} =(|\Zs \Za \rangle \langle \Za \Zs|+|\Js \Ja\rangle
\langle \Ja \Js |)/2 \eqno(9b) $$
results in zero discord, but only when the preferred basis
${\s} = \{ |0 \rangle, |1 \rangle \} $ is employed. The entangled state
of Eq. (9a) could be converted into the mixture of Eq. (9b) through
einselection of the preferred (pointer) basis$^{4,8-11}$ or -- and this
is why decoherence can be regarded as monitoring by the environment --
through a measurement with an undisclosed outcome carried out in the same
pointer basis ${\s} = \{ |0 \rangle, ~ |1 \rangle \} $.

In general, the ignorance of the outsider cannot decrease (but may increase)
as a result of a measurement of a known observable (by the insider), as
the outsider does not know the outcome$^{12}$. Hence,
$$ \delta(\S| \A_{\s}) = H_{\A}(\S, \A_{\s}) - H(\S, \A) \geq 0 \eqno(10)$$
Discord disappears only when $\rho_{\S\A}$ remains unaffected by a partial
measurement of $\s$ on the $\A$ end of the pair -- when the information
is locally accessible.

\section{Demons and discord}

The relevance of the discord for the performance of Maxwell's demon can be now
appreciated. Demons are insiders. They use the acquired information to extract
work from their surroundings. The traditional scenario starts with an
interaction establishing initial correlation between the system and
the apparatus. The demon then
reads off the state of $\A$, and uses so acquired information about $\S$
to extract work by letting $\S$ expand throughout the available phase
(or Hilbert) space of volume (dimension) $d_{\S}$ while in contact with
the thermal reservoir at temperature $T$$^{1,2,13-19}$. This yields:
$$ W^+ = k_{B_2} T (\lg d_{\S} - H(\S|\A)) \eqno(11) $$
of work obtained at a price:
$$ W^- = k_{B_2} T H(\A) \eqno(12) $$
Above, $k_{B_2}$ is the Boltzmann constant adapted to deal with the entropy
expressed in bits and $T$ is the temperature of the heat bath.
The net gain is then:
$$ W = k_{B_2} T (\lg d_{\S} - [H(\A) + H(\S|\A)]) \eqno(13a)$$
The price $W^-$ is the cost of restoring the apparatus to the initial
ready-to-measure state. The significance of this ``cost of erasure'' for
the second law was pointed out in the seminal paper of Szilard$^{13}$. Its
relevance in the context of information processing was elucidated and codified
by Landauer$^{14}$.

It is now accepted that, because of the cost of erasure, neither
classical$^{15-17}$ nor quantum$^{18-21}$ demons can violate the second law.
However, a demon with a supply of empty memory (used to store measurement
outcomes) can extract, on the average, $W^+$ of work per step from a thermal
reservoir. This strategy works, because, in effect, demon is using its empty
memory as a zero entropy (and, hence, $T=0$) reservoir: A memory block of
size $d_{\A}$ is used up with each new measurement. This is expensive
(and wasteful) and only fraudulent accounting (uncovered by Szilard
and Landauer)
that ignores thermodynamic cost of empty memory can create appearance
of a violation of the second law.

To optimize performance demon should use memory of $\A$ more efficiently.
The obvious strategy here is to compress bits of the data after a sequence of
measurements, freeing an unused block of $\Delta \mu$ bits. Demon can compress
data $A_k$ to the size given by $K(A_k)$, their algorithmic complexity$^{22}$.
With compression $ \Delta \mu = \lg d_{\A} - K(A_k) $ memory bits per cycle
are saved. Moreover, one can show that for long sequences of data
the approximate equality $ \langle K(A_k) \rangle \simeq H(\A) $
becomes exact, so that the saved up memory is on the average
$ \Delta \mu = \lg d_{\A} - H(\A) $. By being frugal, classical
Maxwell's demon can gain, per step,
net work of$^{17,21}$:
$$ W = k_{B_2} T (\lg d_{\S} d_{\A} - [ H(\A) + H(\S | \A)]) \eqno(13b) $$
When $\S$ and $\A$ are classically correlated so that Eq. (2) applies, this
can be written as:
$$ W = k_{B_2} T (\lg d_{\S} d_{\A} - H(\S , \A)) \eqno(13c) $$
We note that the efficiency is ultimately determined by the information about
$\S$ and $\A$ {\it accessible} to the demon, and that the same equation would
have followed if we simply regarded the $\S\A$ pair as a composite system, and
the demon used it all up as a fuel.

The efficiency of demons is then determined by the accessible information
about the pair $\S\A$ -- the relevant joint entropy -- and we have already seen
in Eqs. (7) and (8)
that in quantum physics it depends on how the information about the pair is
acquired. A classical demon is local -- it operates one-system-at-a-time
on the correlated quantum pair $\S \A$. In this case the above sketch of the
``standard operating procedure" applies with one obvious {\it caveat}: It needs
to be completed by the specification of the basis demon measures in $\A$.
The cost of erasure is still given by Eq. (12), also for classical demons
extracting work from quantum systems$^{11-13}$, although the relevant $H(\A)$
may increase as a result of decoherence that converts quantum entanglement
into classical data$^{23}$. Thus classical demons operating on pairs of quantum
systems gain net work of:
$$ W^C/k_{B_2}T = \lg d_{\S \A} -  [H(\A) + H(\S|\A)]_{\s} \eqno(14)$$
The only difference between the classical Eq. (13a) and the quantum Eq. (14) is
the obvious dependence on the basis $\s$ demon selects to measure.
The expression in square brackets is the measure of the remaining
(conditional) ignorance and of the cost of erasure. We shall be interested in
the ${\s}$ that maximize $W^C$.

A quantum demon can typically extract more work -- get away with lower costs
of erasure -- because its measurement can be carried out in a global basis
in the combined Hilbert space of $\S\A$ corresponding to observables that
commute with the initial $\rho_{\S\A}$ and avoid increase of entropy associated
with decoherence$^{4,8-11,23}$. The work that can be extracted after
the apparatus gets reset to its ready-to-measure state is:
$$ W^Q/k_{B_2} T = \lg d_{\S \A} - H(\S , \A) \eqno(15)$$
The other way to arrive at Eq. (15) is to use quantum demon in
its capacity of a universal quantum computer, which, by definition, can
transform any state in the Hilbert space into any other state (see Fig. 2 for
an example of a model demon that operates on pairs of qubits). This allows
the quantum demon to reversibly evolve entangled eigenstates of an arbitrary
known $\rho_{\S \A}$ into product states of some $\tilde \rho_{\S\A}$ with the
same eigenvalues, and, hence, same entropy which can be then manipulated
in a local basis that does not perturb its eigenstates, and, hence,
as viewed by the outsider, it will not suffer any additional increase of
entropy. The work extracted by the optimal quantum demon is limited simply
by the basis-independent joint von Neumann entropy of the initial
$\rho_{\S\A}$,
Eq. (8).

The difference between the efficiency of the quantum and classical demons
can be now immediately characterized:
$$ \Delta =\Delta W/k_{B_2} T =  [H(\A) + H(\S|\A)]_{\s} - H(\S ,\A)
\eqno(16)$$
or:
$$\Delta W = k_{B_2} T \delta(\S|\A_{\s}) \eqno(17)$$
Equation (17) relating the extra work $\Delta W = W^Q - W^C$ to quantum discord
-- to the difference of the accessible joint entropy of classical (local)
and quantum (global) demons -- is the principal result of our paper. It answers
an interesting ``demonic" question while simultaneously providing an
operational interpretation of discord.

To gain further insight into implications of the above, let us
first note that discord is, in general, basis dependent. Discord disappears iff
the density matrix has the ``post-decoherence" (or ``post-measurement'') form,
Eq. (6), {\it already before the measurement}.
Given the ability of classical demons to match quantum performance standard
in this case, basis $\s$ that allows for the disappearance of discord in the
presence of non-trivial correlation can be deemed classical$^{3-5}$. We note
that $\rho_{\S\A}$ in the locally diagonal form, Eq. (6), may emerge
as a consequence of the coupling of $\A$ with the environment$^{8-11}$.
The preferred {\it pointer basis} is a result of einselection.

A typical $\rho_{\S\A}$ does not have the form of Eq. (6), however. In that
case discord does not completely disappear for any basis. The least discord:
$$ \hat \delta(\S|\A) = min_{\s}[H(\A) + H(\S | \A)]_{\s} - H(\S,\A)
\eqno(18)$$
corresponds to maximum efficiency of a classical demon. Note that to get the
right answer we had to minimize the sum of the two terms contributing to
the joint entropy, rather than each of them separately. The alternative
$ \hat \partial(\S|\A) = H(\A) + min_{\s} H(\S | \A)_{\s}-H(\S\A) $ would have
followed if the cancellation in Eq. (3) was carried out. The difference between
them is obvious, and $\hat \delta(\S|\A) \geq \hat \partial(\S|\A)$.

\medskip
\section{Demons and discord}

Discord is not symmetric between the two ends of the correlation: In general,
$\hat \delta(\S|\A) \neq \hat \delta(\A|\S) $.
In particular, for density matrices that emerge following einselection in $\A$
$\hat \delta(\S|\A)$ will vanish but $\hat \delta(\A|\S)$ may remain finite.
Such locally accessible correlations are {\it one-way classical}. They are
characterized by a preferred direction -- from $\A$ to $\S$ -- in which more
information about the joint state can be accessed. Thus, when a local demon
can choose between the two ``ends'' of the $\S\A$ pair, it may be more
efficient than a one-way demon. Indeed, one could define polarization:
$$ \varpi(\S|\A) = \hat \delta(\S|\A) - \hat \delta(\A|\S) \eqno(19)$$
to quantify this directionality.

One can generalize discord to collections of correlated quantum systems.
By analogy with the case of a single pair we define it as a difference
between the joint entropy accessible through a particular sequence of (possibly
conditional) measurements -- that is, the obvious generalization of Eq. (7)
-- and the joint von Neumann entropy of the unmeasured density matrix.
The least discord of such a collection of systems is a minimum over all
possible sequences of all possible measurements. This corresponds
to the demon having a choice of the end of the pair it can measure first.

This last situation allows one to address questions raised in the recent
paper on the work that can be extracted by local and global observers from
correlated pairs of quantum systems${^{24}}$. The authors show that a global
observer will be able to extract more work from a pair of quantum systems
than ``Alice and Bob", who can carry out local operations and communicate
classically (``LOCC") with each other, and the difference $\Delta$
(in units of $k_{B_2} T$) is bounded from below:
$$\Delta\geq \check \Delta = max_{\cal A,S}(H({\cal A}),H({\cal
S}))~-~H({\cal A,S})
\eqno(20)$$
This lower bound is illustrated in Fig. 3 along with our result given by
$\hat \delta$ for Werner states. In this case discord yields the work deficit,
$\Delta = \hat \delta$. Indeed, our arguments throughout the paper show that
$\hat \delta$ gives the difference in efficiencies when only one-way classical
communication is allowed. Obviously, allowing for a two-way communication
between Alice and Bob can only help, so $\hat \delta \geq \Delta$ is an upper
bound on the difference of the efficiencies, and there are cases (e.g., Werner
states of Fig. 3) where this upper bound is saturated.

This leads to an interesting questions: When does two-way communication provide
a significant advantage? Does a single round of two-way communication always
suffice, or is it possible that many iterations may help even more? The nested
density matrix of the form:
$$\rho_{\cal A,B,C,...,S}=\sum_i p_{\cal A}(i)\rho_{{\cal
B,C,...,S}|A_i\rangle}
\otimes |A_i\rangle \langle A_i | \eqno(21a)$$
$$\rho_{{\cal B,C,...,S}|A_i\rangle}=\sum_j p_{\cal B}^{(i)}(j)
\rho_{{\cal C,...,S}|A_i,B_j^{(i)}\rangle} \otimes |B^{(i)}_j\rangle
\langle B^{(i)}_j|
   \ ... \ etc.\eqno(21b)$$
where every second of the subsystems (i.e., ${\cal A,C,E,...}$ is on Alice's
side, while the complement ${\cal B,D,...}$ are on Bob's side) settles these
questions.
When the conditional density matrices $ \rho_{{\cal B,C,...,S}|A_i\rangle}$,
$\rho_{{\cal C,...,S}|A_i,B_j\rangle}$, etc., are not co-diagonal in the
relevant Hilbert spaces describing ${\cal B, C, D,...}$ Alice and Bob will
have to exchange data after each measurement to decide what to measure next
if they are to extract all of the potentially accessible information. This is
an example of a situation where a number of back-and-forth
exchanges equal to the number of ``nestings" is necessary to extract all of
the work -- to access all of the locally accessible information. It is tempting
to suggest that such nested density matrices could be used to restrict access
to information by hiding it in some sufficiently deep layer (e. g.,
${\cal S}$),
accessible only if two (or many) parties cooperate in its retrieval.

\medskip
\section{Summary}

First hints of the quantum underpinnings of the Universe emerged over
a century ago in a thermodynamic setting involving black body radiation.
We have studied here implications of quantum physics -- and, in particular,
of the quantum aspects of correlations -- for
classical and quantum Maxwell's demons. We have seen
that discord is a measure of the advantage afforded by the quantum conditional
dynamics, and shown how this advantage is eliminated by decoherence
and the ensuing einselection. Our discussion sheds a new light on the
problem of
transition between quantum and classical: It leads to an operational measure
of the quantum aspect of correlations. As was already pointed out$^{3-5}$,
the aspect of quantumness captured by discord is not the entanglement. Rather,
it is related to the degree to which quantum superpositions are implicated
in a state of a pair (or of a collection) of quantum systems. We expect it
to be relevant in questions involving quantum theory and thermodynamics,
but discord may be also of use in characterizing multiply
correlated states that find applications in quantum computation.

This research was supported in part by the National Security Agency.
Stimulating exchanges of ideas with Harold Ollivier, David Poulin, and
the Horodecki family are gratefully acknowledged.
\vfill
\eject


\noindent{\bf Figure Captions}

\begin{figure}
\epsfig{file=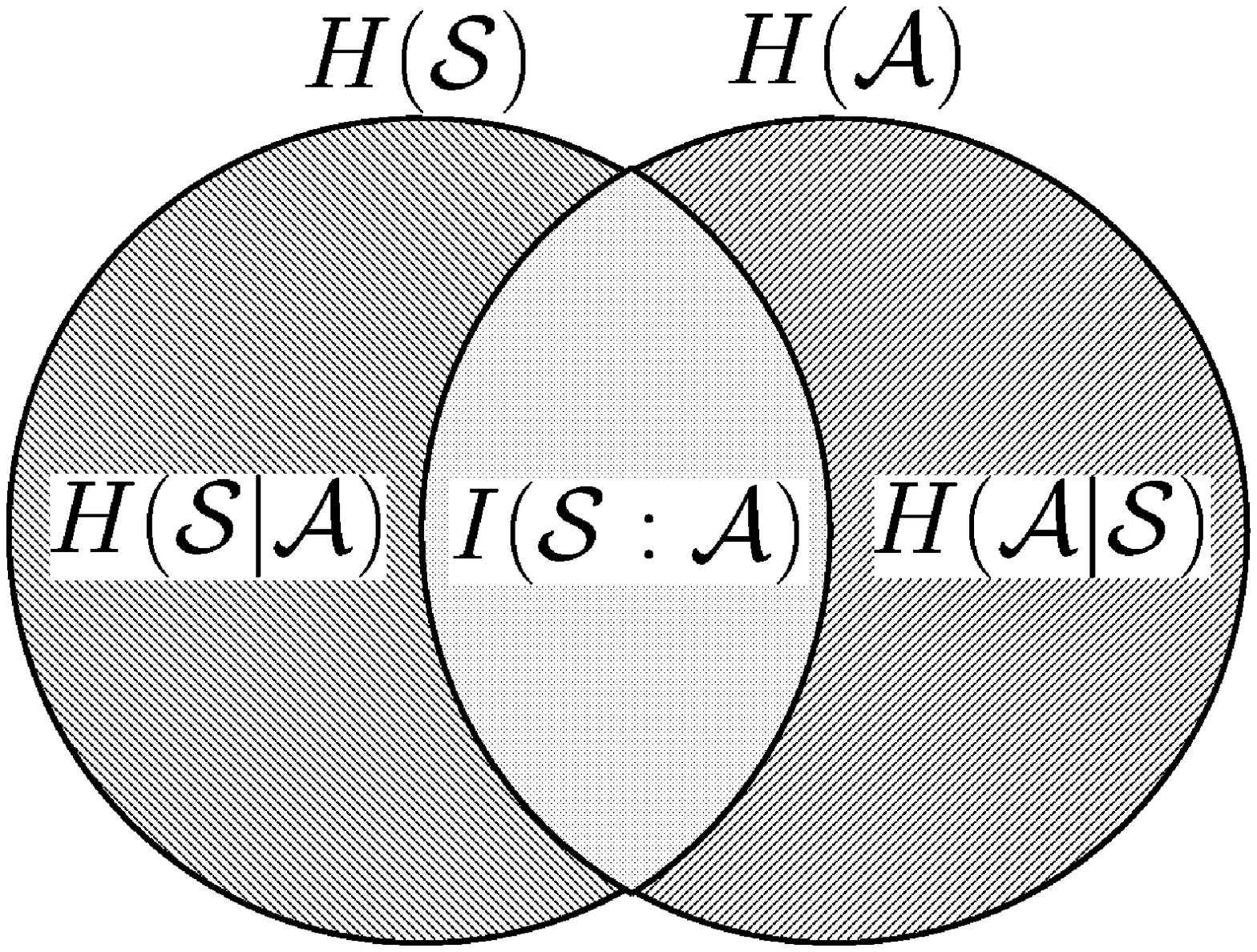,width=5.5cm}
\vspace{.5 in}

\noindent{\bf Fig. 1} Information - theoretic measures of the relationship
between ${\cal A}$ and ${\cal S}$ can be illustrated by means of the
{\it Venn diagram} shown above. Shaded areas represent various uncertainties.
Joint entropy $H({\cal S,A})$ is the
measure of uncertainty about the combined state of ${\cal S}$ and ${\cal A}$.
Individual circles correspond to the uncertainty about ${\cal S}$
and ${\cal A}$. When their states are correlated, the two circles overlapp.
Mutual information $I({\cal S : A})$ is the area of that overlapp. Conditional
entropy $H({\cal S} | {\cal A})$ is one of the half-moons above -- the
one left when the lens corresponding to the mutual information
$I({\cal S} : {\cal A})$ is subtracted from $H({\cal S})$. Obviously,
$H({\cal S, A}) = H({\cal A}) + H({\cal S}|{\cal A}) = H({\cal S})
+ H({\cal A}|{\cal S}) = H({\cal S}) + H({\cal A}) - I ({\cal S}:{\cal A})$.
These equalities are predicated on the classical assumption that the states
of ${\cal S}$ and ${\cal A}$ exist objectively, and, thus, a measurement
need not disturb them. In quantum theory this is not the case:
a measurement will in general redefine the state of the measured object,
even for a ``outsider" who does not know its outcome. Indeed, for a generic
quantum state of the pair ${\cal SA}$ a measurement of ${\cal A}$ alone
would increase the uncertainty of the outsider -- would increase the entropy
he attributes to the pair. This is a consequence of the difference between
the nature of joint states in classical physics (where they are represented
by Cartesian products of subspaces of the constituents) and quantum physics
(where they exist in a tensor product of the two Hilbert spaces). It has
profound effects on the accessibility of the information and leads to
a difference in the efficiency of Maxwell's demons.
\end{figure}
\medskip

\begin{figure}
\epsfig{file=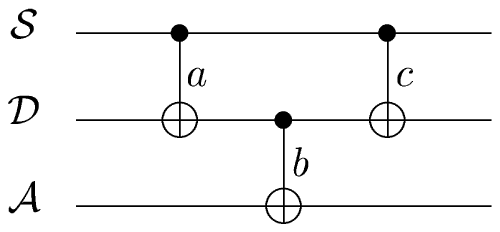,width=7.5cm}
\vspace{.5 in}

\noindent{\bf Fig. 2} A simple model of information processing built of
the controlled not ({\tt c-not}) logical gates illustrates the origin of
the difference between the efficiencies of the classical and quantum versions
of Maxwell's demon. The aim of this demon (whose memory is marked by
${\cal D}$)
is to use the correlation between ${\cal S}$ and ${\cal A}$ to purify their
individual states, so that they can be used individually as fuel. To this end,
the demon; (a) finds out the state of the apparatus ${\cal A}$;
(b) uses this information about ${\cal A}$ and the ``promise"
of the correlation -- prior knowledge about the form of $\rho_{\cal SA}$
embodied in the logical circuit above -- to decrease entropy of ${\cal S}$;
so that $W^+ = k_{B_2} T (1-H({\cal S} | {\cal A}))$ work can be extracted from
${\cal S}$, (c) resets his own memory to the initial ready - to - measure
state, so that the same sequence of actions can be carried out cyclically
on a whole ensemble of identical ${\cal SA}$ pairs. Thus, a quantum demon
operating on an initially entangled state will result in an evolution:
$$ (\a|\Zs \Za \rangle + \b |\Js \Ja \rangle)|\0d \rangle =
\hspace{1.7 in}$$ $$
\a|\Zs \0d \Za \rangle + \b |\Js \0d \Ja \rangle \Rightarrow^a
\a|\Zs \0d \Za \rangle + \b |\Js \1d \Ja \rangle $$
$$\Rightarrow^b  \a|\Zs \0d \Za \rangle + \b |\Zs \1d \Ja \rangle \Rightarrow^c
\a|\Zs \0d \Za \rangle + \b |\Zs \0d \Ja \rangle $$ $$ =
|\Zs\rangle |\0d \rangle (\a|\Za \rangle + \b |\Ja \rangle) \hspace{1.7 in}$$
disentangling ${\cal S}$ from ${\cal A}$. However, a demon whose
memory decoheres -- e.g., entangles with the environment -- will
not be able to take advantage of the quantum correlations in the
state of ${\cal SA}$ pair. Decoherence leading to the einselection
of the basis $\{|\0d \rangle, |\1d \rangle\}$ in the memory of the
demon can be represented
by another {\tt c-not} that acts between ${\cal D}$ influencing the state
of the environment ${\cal E}$ (not shown in the figure). As a consequence,
following the {\tt c-not} (a) interaction with the environment leads to
$(\a|\Zs \0d \Za \rangle + \b |\Zs \1d \Ja \rangle) |\varepsilon_0\rangle
\Rightarrow \a|\Zs \0d \Za \rangle | \varepsilon_0\rangle
+ \b |\Zs \1d \Ja \rangle | \varepsilon_1\rangle$.
Thus, when all the other {\tt cnot}'s are carried out;
$|\Zs\rangle |\0d \rangle (\a|\Za \rangle | \varepsilon_0 \rangle +
\b |\Ja \rangle | \varepsilon_1\rangle) $
obtains, leading to the same pure states of the system and the demon, but
(in effect) a mixed state of the apparatus, $$\rho_{\A}=Tr_{\cal E}
(\a|\Za\rangle |\varepsilon_0 \rangle + \b |\Ja \rangle | \varepsilon_1\rangle)
(\a\langle\Za |\langle\varepsilon_0|+\b \langle \Ja | \langle
\varepsilon_1|) $$ $$
= |\a|^2|\Za\rangle\langle\Za| + |\b|^2|\Ja\rangle \langle \Ja|
\hspace{1.7 in}$$
providing that $\langle \varepsilon_0|\varepsilon_1 \rangle=0$.
In this case, decoherence which turns demons performance from quantum
to classical makes it impossible to extract all of the thermodynamic benefit
from quantum correlations. We leave it as an exercise for the reader to show
that the classical correlation between ${\S}$ and ${\cal A}$ (Eq. (9b))
leads to the same final state, thus proving that a classical demon can
extract all of the work present in a classical correlation.
\end{figure}

\medskip
\begin{figure}
\epsfig{file=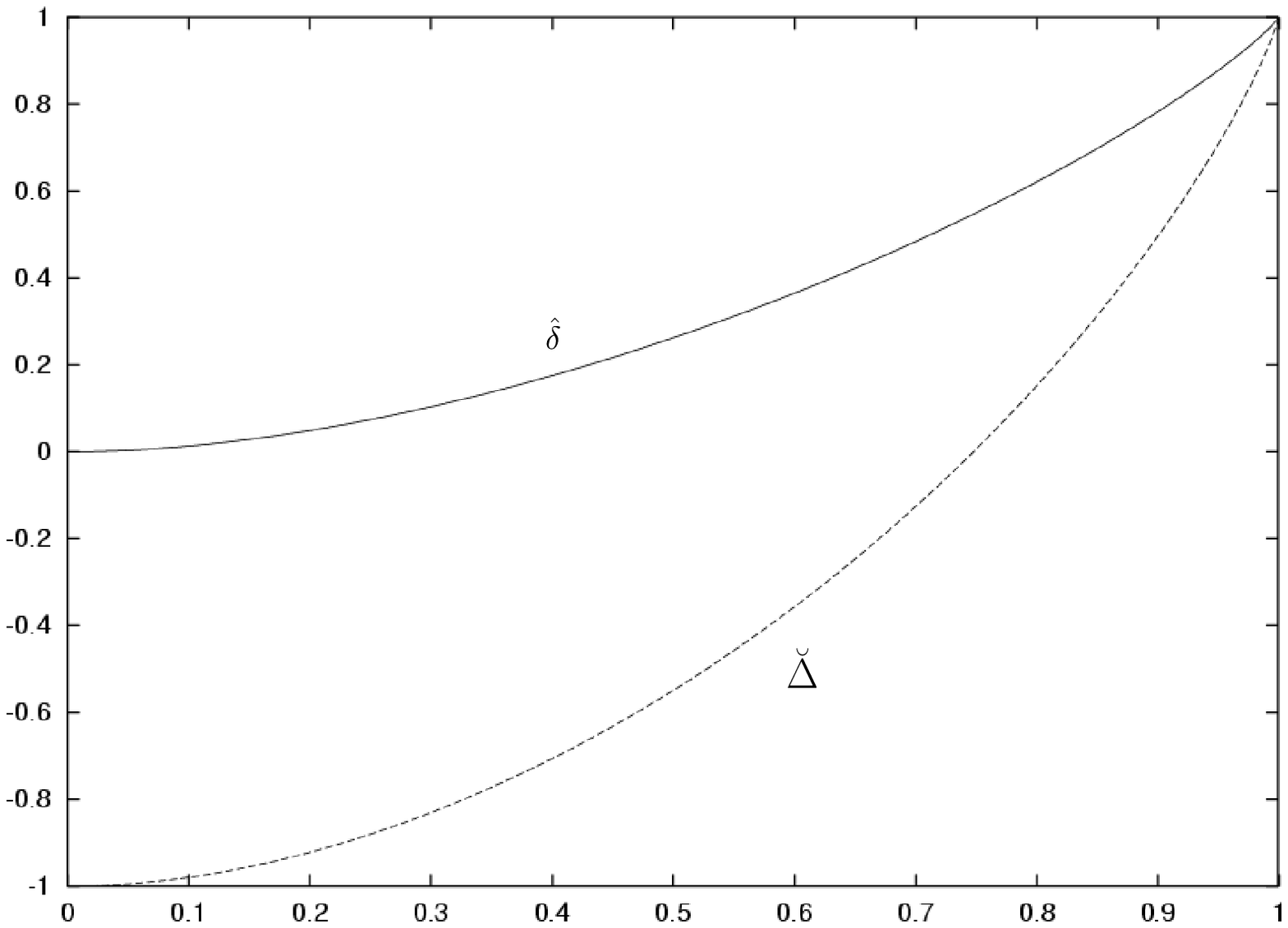,width=7.5cm}
\vspace{.5 in}

\noindent{\bf Fig. 3} Discord $\hat \delta(z)$, Eq. (17), and the lower bound
on the work deficit $\check \Delta(z)$ for Werner states,
$\rho_{\cal SA}={{1-z}\over 4}
{\bf 1} + z |\psi_{\cal SA} \rangle \langle \psi_{\cal SA} |$, where
$ |\psi_{\cal SA} \rangle = (|\Zs \Za\rangle + |\Js \Ja\rangle)/\sqrt 2 $.
In this simple case both
discord (which is equal to the work deficit) and the lower bound on the work
deficit derived in Ref. 24 are independent of the basis and the same for both
``ends" of the correlated pair. As argued here, there are cases
where discord will actually play a role of an upper bound on the work deficit,
as it is derived under the assumption of one-way classical communication.
\end{figure}
\end{document}